\documentclass{tlp}
\usepackage{times}
\usepackage{helvet}
\usepackage{courier}
\usepackage{amsmath}
\usepackage{amssymb}
\usepackage{fancyhdr}
\usepackage{url}

\def\o{\overline}

\def\ar{\leftarrow}
\def\rar{\rightarrow}
\def\lrar{\leftrightarrow}
\def\beq{\begin{equation}}
\def\eeq#1{\label{#1}\end{equation}}
\def\ba{\begin{array}}
\def\ea{\end{array}}
\def\i#1{\hbox{\it #1\/}}

\def\no{{\it not}}

\newtheorem{theorem}{Theorem}

\title{\bf Abstract Gringo}
\author[Gebser et al.]
{MARTIN GEBSER\thanks{Supported by AoF (grant
251170) and DFG (grants SCHA 550/8 and 550/9).}
\\
Aalto University, HIIT, Finland\\
University of Potsdam, Germany\\
gebser@cs.uni-potsdam.de\\
\and
AMELIA HARRISON\thanks{Partially supported
by the National Science Foundation under Grant IIS-1422455.} \\
Univeristy of Texas at Austin, USA\\
ameliaj@cs.utexas.edu\\
\and
ROLAND KAMINSKI$\highast$\\
University of Potsdam, Germany\\
kaminski@cs.uni-potsdam.de\\
\and
VLADIMIR LIFSCHITZ$\dagger$\\
Univeristy of Texas at Austin, USA\\
vl@cs.utexas.edu\\
\and
TORSTEN SCHAUB$\highast$\thanks{Affiliated with Simon Fraser
University, Canada, and IIIS
Griffith University, Australia.}\\
University of Potsdam, Germany\\
INRIA Rennes, France\\
torsten@cs.uni-potsdam.de\\
}

\begin{document}
\maketitle
\pagenumbering{gobble}

\begin{abstract}
This paper defines the syntax and semantics of the input language of the
ASP grounder {\sc gringo}.  The definition covers several constructs
that were not discussed in earlier work on the semantics of that language,
including intervals, pools,
division of integers, aggregates with non-numeric values,
and lparse-style aggregate expressions.  The definition is abstract in the
sense that it disregards some details related to representing programs by
strings of ASCII characters.
It serves as a specification for {\sc gringo} from Version 4.5 on.

This is a corrected version of the paper published in Theory and Practice of
Logic Programming, Volume 15, Issue
04-05 (the special issue on the 31st International Conference on Logic
Programming).
\end{abstract}

\section{Introduction}

Version~4.0 of the ASP grounder {\sc gringo} was released in March of
2013.\footnote{\tt http://potassco.sourceforge.net/ \label{ft1}}
\citeANP{har14a}
\citeyear{har14a} defined the semantics of a subset of its input language
in terms of stable models of infinitary propositional formulas
\cite{tru12}.

That subset does not include, however, several constructs that are frequently
used in ASP programs.  One such construct is integer intervals.
Take, for instance, the ASP solution
to the $n$-queens problem shown in Table~\ref{Kappa}. (It is similar to one of
the solutions in the language of Version~3 presented by \citeANP{geb11b}
\citeyear{geb11b}.)
\begin{table}\label{Kappa}
\begin{tabular}{l}
  \hline
  \\
  {\tt \% place queens on the chess board}\\
  {\tt \{ q(1..n,1..n) \}.}\\
  \\
  {\tt \% exactly 1 queen per row/column}\\
  {\tt :- X = 1..n, not \#count\{ Y : q(X,Y) \} = 1. }\\
  {\tt :- Y = 1..n, not \#count\{ X : q(X,Y) \} = 1. }\\
  \\
  {\tt \% pre-calculate the diagonals}\\
  {\tt d1(X,Y,X-Y+n) :- X = 1..n, Y = 1..n. }\\
  {\tt d2(X,Y,X+Y-1) :- X = 1..n, Y = 1..n. }\\
  \\
  {\tt \% at most one queen per diagonal}\\
  {\tt :- D = 1..n*2-1, 2 \{ q(X,Y) : d1(X,Y,D) \}. }\\
  {\tt :- D = 1..n*2-1, 2 \{ q(X,Y) : d2(X,Y,D) \}. }\\
  \\
  \hline
\end{tabular}
\caption{An ASP solution to the $n$-queens problem.}
\end{table}
Intervals are used in each rule of this program.  To include intervals, we
have to modify the semantics from \citeANP{har14a} \citeyear{har14a} in
two ways.  First,
we have to say that an arithmetic term denotes, generally, a finite set
of integers, not a single integer.   (And it is not necessarily a set
of consecutive integers, because the language of {\sc gringo} allows us to
write {\tt (1..3)*2}, for instance.  This expression denotes the set
$\{2,4,6\}$.)  Second, in the presence of intervals we cannot treat a choice
rule  {\tt \{A\}} as shorthand for the disjunctive rule
{\tt A $\!;\!$ not$\!$ A}
as proposed by \citeANP{fer05b} \citeyear{fer05b}.  Indeed, the first rule of
the program in
Table~\ref{Kappa} has $2^{n^2}$ stable models; the rule
$$\hbox{\tt q(1..n,1..n) ; not  q(1..n,1..n)}$$
has only 2 stable models.

Another feature of {\sc gringo} not covered by \citeANP{har14a}
\citeyear{har14a}, which is somewhat
similar to integer intervals, is pooling.  Pooling is used, for instance, in
the head of the rule
$$\hbox{\tt p(X;Y) :- q(X,Y)}.$$
(Note that a semicolon, not a comma, separates {\tt X} from {\tt Y} in
the head.) This rule has the same meaning as the pair of rules
$$\ba c
\hbox{\tt p(X) :- q(X,Y).}\\
\hbox{\tt p(Y) :- q(X,Y).}
\ea
$$
Pooling is often used to abbreviate a set of facts.  For instance, instead of
$$\hbox{\tt p(a,5).}\quad\hbox{\tt p(b,10).}\quad\hbox{\tt p(c,12).}$$
we can write
$\hbox{\tt p(a,5;b,10;c,12).}$
In this paper, we talk about ``pools''---groups of terms such as
\hbox{\tt a,5;b,10;c,12.}

Yet another limitation of the proposal from \citeANP{har14a} \citeyear{har14a} is related to the difference between
``dlv-style'' aggregates, such as
\beq
\hbox{\tt not \#count\{ Y : q(X,Y) \} = 1}
\eeq{dlv}
in the second rule of the program shown in Table~\ref{Kappa}, and
``lparse-style'' aggregates, such as
\beq
\hbox{\tt 2 \{ q(X,Y) : d2(X,Y,D) \}}
\eeq{lparse}
in the last rule of the program.  Both expressions have to do with counting.
Syntactically, the difference is that in expression~(\ref{dlv}) both the name
of the aggregate ({\tt \#count}) and the binary relation applied to the result
of counting and a constant ({\tt =}) are shown explicitly; in~(\ref{lparse}),
the fact that the constant~{\tt 2}
occurs on the left, in the lower bound position, tells us that the
relation $\leq$ is applied to that number and to the result of counting.
More importantly, there is a difference between the kinds of objects that we
count.  In case of expression~(\ref{dlv}) we count, for a given value of
{\tt X},  the values of the variable~{\tt Y} such that {\tt q(X,Y)}
belongs to the stable model.  In case of~(\ref{lparse}) we count, for a given
value of~{\tt D},  the atoms {\tt q(X,Y)} that belong to the stable model
and satisfy an additional condition: {\tt d2(X,Y,D)} belongs to the model
as well.  Thus the atom in front of the colon in~(\ref{lparse}) plays two
roles: it tells us what to count, and it gives a condition on the stable
model that needs to be checked.

The language studied by \citeANP{har14a} \citeyear{har14a} does not include
lparse-style
aggregates.  The easiest way to add such aggregates is to treat them as
abbreviations.  For instance,~(\ref{lparse}) can be viewed as shorthand for
the dlv-style expression
$$\hbox{\tt 2 <= \#count\{ q(X,Y) : q(X,Y), d2(X,Y,D) \}}.$$
(In this paper we adopt a more elaborate translation that allows us to
accommodate negated atoms in front of the colon.)  In this expression, the
first occurrence of {\tt q(X,Y)} is syntactically a term, and the second is
an atom.  Thus treating~(\ref{lparse}) as an abbreviation depends on the
possibility of using the same symbol as a function and as a predicate.  This is
customary in Prolog, but not in first-order logic, and this was not allowed by
\citeANP{har14a} \citeyear{har14a}.

Our goal is to define the syntax and semantics of the language AG
(short for {\sl Abstract Gringo\/})---a large subset of the input language
of {\sc gringo} that includes the features mentioned previously
and a few other constructs not described by \citeANP{har14a} \citeyear{har14a}.
This is
similar to the work that has led to the definition of the ASP
Core language \cite{aspcore2}.\footnote{Syntactically, AG is essentially
an extension of ASP Core.  But it does not include extra-logical
constructs, such as weak constraints and queries.  The semantics of
aggregates in AG is based on the approach of \citeANP{fer05} \citeyear{fer05}
and thus is not
equivalent to the semantics of aggregates in ASP Core when aggregates
are used recursively in the presence of negation.  Among the language
constructs that are not in
ASP Core, in AG we find pooling, intervals, and conditional literals.
(These constructs originally appeared in the input language of {\sc lparse},
but in AG they are more general; for instance, interval bounds may
contain variables,
and restrictions to ``domain predicates'' have disappeared.)  Unlike
ASP Core, AG supports aggregates in rule heads; see
Section~\ref{abb}.}$^,$\footnote{The definition of ASP Core does not refer to
infinitary objects, such as infinitary propositional formulas used in this
paper.  But it appears that infinitary objects of some kind will be required
to correct the oversight in \cite[Section~2.2]{aspcore2}---the set
\hbox{inst($\{e_1;\dots;e_n\}$)}, included in the body of a rule in the
process of instantiation, can be infinite.}

The semantics of AG defined in this paper serves a specification for
{\sc gringo} from Version~4.5 on.  It can be used to prove
the correctness of programs written in its input language.
As an example, in the electronic appendix we prove the
correctness of the program shown in Table~\ref{Kappa}.

AG is abstract in the sense that its definition disregards
some details related to representing programs by strings of ASCII characters.
For example, semicolons are used in the input language of {\sc gringo} in
at least
three ways: to denote disjunction in the head of a rule, conjunction in the
body, and pooling within an atom.  In AG, three different symbols play
these different roles.  The richer alphabet of AG makes it easier to define the
semantics of the language and to reason about ASP programs.


\section{Syntax of AG} \label{sec:syntax}

\subsection{Symbols and Terms} \label{ssec:symbols}

We assume that five sets of symbols are selected:
{\sl numerals}, {\sl symbolic constants}, {\sl negated constants},
{\sl variables}, and {\sl aggregate names}. We assume that a 1--1
correspondence between the set of symbolic constants and the set of negated
constants is chosen.  For every symbolic constant $p$, the corresponding
negated constant will be called its {\sl strong negation} and denoted by
$\tilde{p}$.
Further, we assume that these sets do not contain the
symbols
\beq
+\qquad -\qquad \times\qquad /\qquad..
\eeq{ops}
\vskip -.5cm
\beq
{\it inf} \qquad {\it sup}
\eeq{exts}
\beq
= \qquad \not = \qquad <\qquad >\qquad \leq\qquad \geq
\eeq{comps}
\beq
\bot\qquad\no\qquad\land\qquad \lor\qquad\ar
\eeq{conns}
\beq
,\qquad ;\qquad :\qquad (\qquad )\qquad \{\qquad \} \qquad \langle \qquad
\rangle
\eeq{punct}
and that they are pairwise disjoint. All these symbols together form the
alphabet of AG, and AG rules will be defined as strings over this alphabet.

When a symbol is represented in ASCII, its type is determined by its first
two characters. For instance, a numeral starts with a digit or~{\tt -}
followed by a digit. A symbolic constant starts with a lower-case letter.
A negated constant starts with~{\tt -} followed by a lower-case letter.
A variable starts with an upper-case letter, and an aggregate name starts
with~{\tt \#}. (The strings {\tt \#false}, {\tt \#inf}, and ~{\tt \#sup},
which represent $\bot$, {\it inf}, and {\it sup}, also  start with~{\tt \#}.)
The symbols $\langle$ and $\rangle$ (which are used to indicate the
boundaries of a tuple within a term) correspond to the ASCII characters
{\tt (} and {\tt )}.\footnote{When an AG term representing a tuple of
length~1, such as $\langle a \rangle$, is represented in ASCII, a comma
is appended to the tuple: {\tt (a,)} .}
Each of the symbols (\ref{ops})--(\ref{punct})
except for $\land$ and~$\lor$ has a unique ASCII representation;
the symbols $\land$ and $\lor$
can be represented by semicolons and in some cases also by commas.

We assume that a 1--1 correspondence between the set of numerals and the
set~${\bf Z}$ of integers is chosen.  For every integer~$n$, the corresponding
numeral will be denoted by~$\overline n$.

{\sl Terms} are defined recursively, as follows:
\begin{itemize}
\item all numerals, symbolic constants, and variables are terms;
\item if~$f$ is a symbolic constant and~${\bf t}$ is a tuple\footnote{
In this paper, when we refer to a {\sl tuple} of syntactic objects, we mean
that the tuple may be empty and that its members are separated by commas.}
of terms then $f({\bf t})$ is a term;
\item if $t_1$ and $t_2$ are  terms and $\star$ is one of the
symbols~(\ref{ops}) then $(t_1\star t_2)$ is a  term;
\item if ${\bf t}$ is a tuple of terms then
$\langle{\bf t}\rangle$ is a term.
\end{itemize}

In a term of the form $f()$ the parentheses can be dropped, so that every
symbolic constant can be viewed as a term.  In a term of the form
$(t_1\star t_2)$ we will drop the parentheses when it should not lead to
confusion. A term of the form $(\overline 0 - t)$ can be abbreviated as $-t$.

A term, or a tuple of terms, is {\sl precomputed} if it contains neither
variables nor symbols~(\ref{ops}).
We assume a total order on precomputed terms such that {\it inf} is its least
element, {\it sup} is its greatest element, and,
for any integers $m$ and $n$, $\overline m \leq \overline n$ iff $m \leq n$.

We assume that for each aggregate name~$\alpha$ a function $\widehat \alpha$
is chosen that maps every set of tuples of precomputed terms to a precomputed
term.\footnote{\label{ft3}This understanding of $\widehat \alpha$ is different from that
given by \citeANP{har14a} \citeyear[Section~3.3]{har14a}. There, $\widehat \alpha$ is understood as
a function that maps tuples of precomputed terms to elements of $\bf Z \cup
\{\infty, - \infty\}$. }
The AG counterparts of the aggregates implemented in Version 4.5 of
{\sc gringo} are defined below using the following terminology.
If the first member of a tuple~$\bf t$
of precomputed terms is a numeral $\overline n$ then we say that the integer~$n$ is
the {\sl weight} of~$\bf t$; if $\bf t$ is empty or its first member is not
an integer then the weight of~$\bf t$ is~0.
For any set~$T$ of tuples of precomputed terms,
\begin{itemize}
\item $\widehat{{\it count}}(T)$ is the numeral corresponding to the cardinality of~$T$ if~$T$ is finite, and
${\it sup}$ otherwise;
\item $\widehat{{\it sum}}(T)$ is the numeral corresponding to the sum of the
weights of all tuples in~$T$
if~$T$ contains finitely many tuples with non-zero weights, and $0$ otherwise;
\item $\widehat{{\it sum+}}(T)$ is the numeral corresponding to the sum of the
weights of all tuples in~$T$
whose weights are positive if~$T$ contains finitely many such tuples,
and {\it sup} otherwise;
\item $\widehat{{\it min}}(T)$ is~${\it sup}$ if~$T$ is empty, the
least element of the set consisting of the first elements of the tuples in $T$
if $T$ is a finite non-empty set, and {\it inf} if~$T$ is infinite;
\item
$\widehat{{\it max}}(T)$ is~{\it inf} if~$T$ is empty, the greatest element
of the set consisting of the first elements of the tuples in $T$
if $T$ is a finite non-empty set, and {\it sup} if~$T$ is infinite.
\end{itemize}

\subsection{Atoms, Literals, and Choice Expressions}

A {\sl pool} is an expression of the form
${\bf t}_1 ; \dots ; {\bf t}_n$  where $n \geq 1$ and each ${\bf t}_i$
is a tuple of terms.\footnote{This form of pooling is less general than what
is allowed in the input language of {\sc gringo}.  For instance, $f(a;b)$ is
neither a term nor a pool.}
In particular, every tuple of terms is a pool.

An {\sl atom} is a string of one of the forms
$p(P),\ \tilde{p}(P)$
where $p$ is a symbolic constant and $P$ is a pool.
In an atom of the form $p()$ or $\tilde{p}()$ the parentheses can be
dropped, so that all symbolic constants and all negated constants can be viewed
as atoms.

For any atom~$A$, the strings
\beq
A\qquad\no\ A\qquad\no\ \no\ A
\eeq{ls}
are {\sl symbolic literals}.\footnote{Semantically, the status of
``double negations'' in AG is the same as in logic programs with
nested expressions \cite{lif99d}, where conjunction, disjunction, and
negation can be nested arbitrarily.
Dropping a double negation may change the meaning of a rule.  For
instance, the one-rule program $p\ar \no\ \no\ p$ has two stable models
$\emptyset$, $\{p\}$ (see Section~\ref{sem:inf}); the latter will disappear
if we drop $\no\ \no$.} An {\sl arithmetic literal} is
a string of the form $t_1\prec t_2$
where~$t_1$,~$t_2$ are terms and $\prec$ is one of the symbols (\ref{comps}).

A {\sl conditional literal} is a string of the form
$H : {\bf L}$
where $H$ is a symbolic or arithmetic literal or the symbol $\bot$ and
${\bf L}$ is a tuple of symbolic or arithmetic literals.
If ${\bf L}$ is empty then we will drop the colon, so that every symbolic or
arithmetic literal
can be viewed as a conditional literal.\footnote{In the input language of
{\sc gringo}, dropping the colon when ${\bf L}$ is empty is required.}

An {\sl aggregate atom} is a string of one of the forms
\begin{align}
       \alpha\{{\bf t}_1 : {\bf L}_1\,;\ldots;\,{\bf t}_n : {\bf L}_n\}&\prec s
\label{ag1}\\
s\prec\;\alpha\{{\bf t}_1 : {\bf L}_1\,;\ldots;\,{\bf t}_n : {\bf L}_n\}&
\label{ag2}\\
s_1\prec_1\alpha\{{\bf t}_1 : {\bf L}_1\,;\ldots;\,{\bf t}_n : {\bf L}_n\}&
                                                              \prec_2s_2
\label{ag3}
\end{align}
($n\geq 0$),
where
\begin{itemize}
\item
$\alpha$ is an aggregate name,
\item
each ${\bf t}_i$ is a tuple of terms,
\item
each ${\bf L}_i$ is a tuple of symbolic or arithmetic literals (if
${\bf L}_i$ is empty and ${\bf t}_i$ is nonempty then the preceding colon may
be dropped),
\item
each of $\prec$, $\prec_1$, $\prec_2$ is one of the
symbols~(\ref{comps}),
\item
each of $s$, $s_1$, $s_2$  is a term.
\end{itemize}
For any aggregate atom~$A$, the strings~(\ref{ls})
are {\sl aggregate literals}.

A {\sl literal} is a conditional literal or an aggregate literal.

A {\sl choice expression} is a string of the form $\{A\}$ where
$A$ is an atom.

\subsection{Rules and Programs} \label{rp}

A {\sl rule} is a string of the form
\beq
H_1\,\lor\,\cdots\,\lor\,H_k \ar B_1\,\land\,\cdots\,\land\,B_m
\eeq{rule3}
or of the form
\beq
C \ar B_1\,\land\,\cdots\,\land\,B_m
\eeq{rule4}
($k,m\geq 0$), where each~$H_i$ is a symbolic or arithmetic
literal,\footnote{In the input language of {\sc gringo}, $H_i$ may be any
conditional literal.}~$C$ is a choice
expression, and each~$B_j$ is a literal.
The expression $B_1 \land \dots \land B_m$ is the
{\sl body} of the rule; $H_1 \lor \dots \lor H_k$ is the {\sl head}
of (\ref{rule3}); $C$ is the {\sl head} of (\ref{rule4}).
If the body of a rule is empty and the head is not then the arrow can be
dropped. 

For instance, here are the first five rules of the program from
Table~\ref{Kappa} written in the syntax of~AG:
$$
\ba {ll}
  R_1: \qquad  & \{\; q(\overline 1\;..\;\overline n,\overline 1\;..\;\overline n)\; \},\\
  R_2: \qquad  & \ar \; X = \overline 1\;..\;\overline n \; \land \; \no \; {\it count}\{ Y : q(X,Y) \}\;
= \; \overline 1,\\
  R_3: \qquad  & \ar \; Y = \overline 1\;..\;\overline n \; \land \; \no \; {\it count}\{ X : q(X,Y) \}\;
= \; \overline 1,\\
  R_4: \qquad  & {\it d1}(X,Y,X-Y+\overline n) \; \ar \; X = \overline 1\;..\;\overline n \; \land \;
Y = \overline 1\;..\;\overline n,\\
  R_5: \qquad  & {\it d2}(X,Y,X+Y-\overline 1) \; \ar \; X = \overline 1\;..\;\overline n \; \land \;
Y = \overline 1\;..\;\overline n.
\ea
$$
The other two rules use abbreviations introduced in the next section.

A {\sl program} is a finite set of rules.

\section{Abbreviations}\label{abb}

Let $C$ be an expression of the form
\beq
s_1\prec_1
\alpha\{{\bf t}_1 : L_1 : {\bf L}_1\,;\ldots;\,{\bf t}_n : L_n : {\bf L}_n\}
\prec_2s_2
\eeq{chexp}
($n \geq 0$),
where
each $L_i$ is a symbolic literal of one of the forms
\beq
p({\bf t})\qquad  \no \; p({\bf t})\qquad   \no \; \no \; p({\bf t})
\eeq{forms}
($p$ is a symbolic or negated constant and ${\bf t}$ is a tuple of terms)
and $\alpha$, ${\bf t}_i$, ${\bf L}_i$, $\prec_1$, $\prec_2$, $s_1$, and $s_2$
are as in the definition of an aggregate atom.
Then a string of the form
\beq
C \ar B_1\,\land\,\cdots\,\land\,B_m
\eeq{choice}
($m\geq 0$), where each~$B_j$ is a literal,
 is shorthand for the set of rules consisting of the rule
\beq
\ar B_1\,\land\,\cdots\,\land\,B_m\,\land\;
  \no\ s_1\prec_1
\alpha\{{\bf t}_1 : L_1 , {\bf L}_1\,;\ldots;\,{\bf t}_n : L_n , {\bf L}_n\}
\prec_2s_2
\eeq{choice2}
and, for each $L_i$ in (\ref{chexp}) such that $L_i$ is an atom, the rule
\beq
\{L_i\} \ar B_1\,\land\,\cdots\,\land\,B_m\,\land\,C_i
\eeq{choice1}
where~$C_i$ is the conjunction of the members
of~${\bf L}_i$.

In both (\ref{choice2}) and (\ref{choice1}), the conjunction sign shown
after~$B_m$ should be dropped if $m=0$; in (\ref{choice1}) it should also
be dropped if $C_i$ is empty.  The parts $s_1\prec_1$ and $\prec_2s_2$ in
(\ref{chexp}) are optional; if one of them is
missing then it is dropped from~(\ref{choice2}) as well; if both are missing
then rule~(\ref{choice2}) is dropped from the set altogether.  If $m=0$ in
(\ref{choice}) then the arrow can be dropped.

The {\sl term representations} of literals (\ref{forms}) are the tuples
$$
\overline 0, p({\bf t})\qquad \overline 1,p({\bf t}) \qquad \overline 2, p({\bf t})
$$
of terms. (Each of them is indeed a tuple of terms, because
$p({\bf t})$ can be viewed as a term.)

Also viewed as an abbreviation is any expression of the form
\beq
s_1\,\{L_1:{\bf L}_1\,;\ldots;\,L_n:{\bf L}_n\}\,s_2
\eeq{cardc}
($n>0$), where $s_1$, $s_2$ are terms,
each~$L_i$ is a symbolic literal  of one of the forms
(\ref{forms}) that does not contain $..\;$, and each~${\bf L}_i$ is a tuple
of symbolic or arithmetic literals.\footnote{To be precise, if~${\bf L}_i$
is empty then the colon after~$L_i$ is dropped.}
Such an expression is understood differently
depending on whether it occurs in the head or the body of a rule.
In the head of a rule,~(\ref{cardc}) is understood as shorthand for
an expression of the form (\ref{chexp}):
\beq
s_1\leq
{\it count}\{{\bf t}_1: L_1 : {\bf L}_1\,;\ldots;\,{\bf t}_n: L_n : {\bf L}_n\}
\leq s_2
\eeq{choicehead}
where ${\bf t}_i$ is the term representation of $L_i$.  If either or
both of the terms~$s_1$,~$s_2$ are missing, the abbreviation is
understood in a similar way. (Note that choice expressions that do not
contain~$..$ are
expressions of the form (\ref{cardc}) where both $s_1$ and $s_2$ are missing,
$n =1$, $L_1$ is of the form $p({\bf t})$, and ${\bf L}_1$ is empty. In this
case, we do not view (\ref{cardc}) as an abbreviation.)

In the body of a rule (\ref{cardc}) is understood as shorthand for the
aggregate atom $$s_1\leq{\it count}\{{\bf t}_1:L_1,{\bf L}_1;\ldots;\,
{\bf t}_n :L_n, {\bf L}_n\}\leq s_2$$
where ${\bf t}_i$ is the term representation of $L_i$.\footnote{If~${\bf L}_i$
is empty then the comma after~$L_i$ in this expression should be dropped.} If
either of the terms~$s_1$,~$s_2$ in (\ref{cardc}) is missing, the
abbreviation is understood in a similar way.

These abbreviations can be used, for instance, to represent the last two
rules of the program from Table~\ref{Kappa} in the syntax of~AG:
$$\ba l
\ar \; D = \overline 1\;..\;\overline n*\overline 2-\overline 1 \; \land \;\overline 2 \; \{ q(X,Y)
: d1(X,Y,D) \},\\
\ar \; D = \overline 1\;..\;\overline n*\overline 2-\overline 1 \; \land \;\overline 2 \; \{ q(X,Y)
: d2(X,Y,D) \}.
\ea
$$
Written out in full, these expressions become
$$
\ba{ll}
R_6: \qquad  & \ar \; D = \overline 1\;..\;\overline n*\overline 2-\overline 1
\;\; \land \;\; \overline 2 \leq {\it count}\{ \overline 0, q(X,Y)
:q(X,Y), d1(X,Y,D) \},\\
R_7: \qquad  & \ar \; D = \overline 1\;..\;\overline n*\overline 2-\overline 1
\;\; \land \;\; \overline 2 \leq {\it count}\{ \overline 0,q(X,Y)
:q(X,Y), d2(X,Y,D) \}.
\ea
$$

\section{Semantics of AG} \label{sec:semantics}

We will define the semantics of AG using a syntactic
transformation~$\tau$.  The function~$\tau$ converts rules
into infinitary formulas formed from atoms
of the form $p({\bf t})$ or $\tilde{p}({\bf t})$, where~$p$ is a symbolic
constant, and~${\bf t}$ is a tuple
of precomputed terms.  Then the stable models of a program
will be defined in terms of stable model semantics of infinitary formulas
in the sense of \citeANP{tru12} \citeyear{tru12}, which is reviewed below.

\subsection{Review: Infinitary Propositional Formulas}\label{sem:inf}

Let $\sigma$ be a propositional signature,
that is, a set of propositional atoms.  The sets
\hbox{$\mathcal{F}_0,\mathcal{F}_1,\ldots$} are defined as follows:
\begin{itemize}
\item $\mathcal{F}_0=\sigma$,
\item $\mathcal{F}_{i+1}$ is obtained from $\mathcal{F}_{i}$ by
adding expressions $\mathcal{H}^\land$ and $\mathcal{H}^\lor$ for all subsets
$\mathcal{H}$ of $\mathcal{F}_i$, and expressions \hbox{$F\rar G$}
for all $F,G\in\mathcal{F}_i$.
\end{itemize}
The elements of $\bigcup^{\infty}_{i=0}\mathcal{F}_i$ are called {\sl
(infinitary) formulas} over $\sigma$.

In an infinitary formula, the symbols $\top$ and $\bot$
are understood as abbreviations
for~$\emptyset^{\land}$ and $\emptyset^{\lor}$ respectively;
$\neg F$ stands for $F\rar\bot$, and $F\lrar G$
stands for \hbox{$(F\rar G)\land(G\rar F)$}.

Subsets of a signature~$\sigma$ will also be called its {\sl interpretations}.
The satisfaction relation between an interpretation and a formula is
defined recursively as follows:
\begin{itemize}
\item For every atom $p$ from $\sigma$, $I\models p$ if $p\in I$.
\item $I\models\mathcal{H}^\land$ if for every formula $F$ in~$\mathcal{H}$,
$I\models F$.
\item $I\models\mathcal{H}^\lor$ if there is a formula $F$ in~$\mathcal{H}$
such that $I\models F$.
\item $I\models F\rar G$ if $I\not\models F$ or $I\models G$.
\end{itemize}
We say that an interpretation satisfies a set $\mathcal{H}$ of formulas,
or is a {\sl model} of $\mathcal{H}$, if it satisfies every formula in
$\mathcal{H}$.  Two sets of formulas are {\sl equivalent} if
they have the same models.

The {\sl reduct} $F^I$ of a formula~$F$ w.r.t. an
interpretation~$I$ is defined as follows:
\begin{itemize}
\item For $p\in \sigma$, $p^I=\bot$ if $I\not\models p$; otherwise $p^I=p$.
\item $(\mathcal{H}^\land)^I=\{G^I\ |\ G\in\mathcal{H}\}^\land$.
\item $(\mathcal{H}^\lor)^I=\{G^I\ |\ G\in\mathcal{H}\}^\lor$.
\item $(G\rar H)^I=\bot$ if $I\not\models G\rar H$; otherwise
\hbox{$(G\rar H)^I=G^I\rar H^I$}.
\end{itemize}
An interpretation~$I$ is a {\sl stable model} of a set $\mathcal{H}$ of
formulas if it is minimal w.r.t.\ set inclusion among the interpretations
satisfying the reducts of all formulas from~$\mathcal{H}$.

For instance, if $I=\emptyset$ then
$$(\neg \neg p \rar p)^I=(\neg \neg p)^I\rar p^I=\bot\rar\bot;$$
if $I=\{p\}$ then
$$(\neg \neg p \rar p)^I=(\neg \neg p)^I\rar p^I=\neg (\neg p)^I\rar p
   =\neg\bot\rar p.$$
In both cases,~$I$ is a minimal model of the reduct.  Consequently,
both~$\emptyset$ and $\{p\}$ are stable models of $\{\neg\neg p\rar p\}$.

\subsection{Semantics of Terms and Pools}

A term is {\sl ground} if it does not contain variables. The definition of
``ground'' for pools, symbolic literals, and arithmetic literals is the same.
Semantically, every ground term~$t$  represents a finite
set of precomputed terms~$[t]$, which is defined recursively:
\begin{itemize}
\item if $t$ is a numeral or a symbolic constant then $[t]$ is
$\{t\}$;
\item if $t$ is $f(t_1, \dots, t_n)$  then $[t]$ is the set of
terms $f(r_1, \dots, r_n)$ for all
$r_1 \in [t_1], \dots,$ $r_n \in [t_n]$;
\item if $t$ is $(t_1 + t_2)$ then $[t]$ is the set
of numerals $\overline {n_1 + n_2}$ for all integers $n_1, n_2$ such that
$\overline {n_1} \in [t_1]$ and $\overline {n_2} \in [t_2]$; similarly when $t$ is
$(t_1 - t_2)$ or $(t_1 \times t_2)$;
\item if $t$ is $(t_1 / t_2)$ then $[t]$ is the set
of numerals $\overline{\lfloor n_1/n_2 \rfloor}$ for all
integers $n_1, n_2$ such that
$\overline {n_1} \in [t_1]$, $\overline {n_2} \in [t_2]$, and $n_2\neq 0$;
\item if $t$ is $(t_1\, ..\, t_2)$ then $[t]$ is the set of
numerals~$\overline m$ for all integers $m$ such that,
for some integers $n_1, n_2,$
$$ \overline{n_1} \in [t_1], \qquad \overline{n_2} \in [t_2], \qquad
n_1 \leq m \leq n_2;$$
\item if $t$ is $\langle t_1, \dots, t_n\rangle$
then $[t]$ is the set of terms
$\langle r_1, \dots, r_n \rangle$ for all $r_1 \in [t_1],\dots,$ $r_n \in [t_n]$.
\end{itemize}
This definition is extended to an arbitrary ground pool~$P$; $[P]$
is a finite set of precomputed tuples:
\begin{itemize}
\item if $P$ is a tuple $t_1, \dots, t_n$ of terms $(n \not = 1)$
then $[P]$ is the set of tuples $r_1, \dots, r_n$ for all
\hbox{$r_1 \in [t_1], \dots, r_n \in [t_n]$};
\item if $P$ is a pool ${\bf t}_1; \dots; {\bf t}_n$ $(n > 1)$ then
$[P]$ is $[{\bf t}_1] \cup \dots \cup [{\bf t}_n]$.
\end{itemize}

For instance, $[\overline 1..\overline n,\overline 1..\overline n]$ is the set
$\{\overline i , \overline j : 1 \leq i,j \leq n \}$.

It is clear that if a ground term $t$ contains neither symbolic
constants nor the symbols~$\langle$ and~$\rangle$ then every element of $[t]$
is a numeral.
If a tuple~${\bf t}$ of ground terms is
precomputed then $[{\bf t}]$ is~$\{{\bf t}\}$.
The set~$[t]$ can be empty. For example,
$[1..0] = [1/0] = [1+a] = \emptyset$.

About a tuple of terms that does not contain .. we say that it is
{\sl interval-free}.
It is clear that if a tuple~${\bf t}$ of
ground terms is interval-free then the cardinality of the set
$[{\bf t}]$ is at most~$1$.

\subsection{Semantics of Arithmetic and Symbolic Literals}

For any ground (symbolic or arithmetic) literal $L$ we will define
two translations,
$\tau_{\land}L$ and $\tau_{\lor}L$. The
specific translation function applied to an occurrence of a symbolic or
arithmetic literal in a rule
depends on the context, as we will see in the following sections.

We will first consider symbolic literals.
For any ground atom $A$,
\begin{itemize}
\item
if $A$ is $p(P)$ then $\tau_{\land}A$ is the conjunction of
atoms $p({\bf t})$ over all tuples $\bf t$ in~$[P]$,
and $\tau_{\lor}A$ is the disjunction of these atoms;
\item
if $A$ is $\tilde{p}(P)$ then $\tau_{\land}A$ is the conjunction of
atoms $\tilde{p}({\bf t})$ over all tuples $\bf t$ in~$[P]$,
and $\tau_{\lor}A$ is the disjunction of these atoms;
\item $\tau_{\land} (\no \; A)$ is $\neg \tau_{\lor} A$, and
 $\tau_{\lor} (\no \; A)$ is $\neg \tau_{\land} A$;
\item $\tau_{\land} (\no \; \no \; A)$ is
$\neg\neg \tau_{\land} A$, and
$\tau_{\lor} (\no \; \no \; A)$ is
$\neg\neg \tau_{\lor} A$.
\end{itemize}
The definitions of  $\tau_{\land}$ and $\tau_{\lor}$ for arithmetic literals
 are as follows:
\begin{itemize}
\item $\tau_{\land} (t_1 \prec t_2)$ is
$\top$ if the relation $\prec$ holds between
the terms $r_1$ and $r_2$ for all $r_1 \in
[t_1]$ and $r_2 \in [t_2]$, and $\bot$
otherwise;
\item $\tau_{\lor} (t_1 \prec t_2)$ is
$\top$ if the relation $\prec$ holds between the terms
$r_1$ and $r_2$ for some $r_1, r_2$
such that $r_1 \in [t_1]$ and $r_2 \in [t_2]$, and $\bot$ otherwise.
\end{itemize}
For instance, $\tau_{\lor}p(\overline 2..\overline 4)$ is
$p(\overline 2) \lor p(\overline 3) \lor p(\overline 4)$,
and $\tau_{\lor}(\overline 2=\overline 2..\overline 4)$ is $\top$.

For any tuple ${\bf L}$ of ground literals,
$\tau_{\lor}{\bf L}$ stands for the conjunction of the formulas
$\tau_{\lor}L$ for all
members $L$ of ${\bf L}$. The expressions $\tau_{\land}\bot$ and
$\tau_{\lor}\bot$ both stand for $\bot$.

It is clear that if $A$ has the form $p({\bf t})$ or $\tilde{p}({\bf t})$,
where ${\bf t}$ is a tuple of precomputed terms, then each of the
formulas $\tau_{\land} A$ and $\tau_{\lor} A$ is $A$.

\subsection{Semantics of Choice Expressions}

The result of applying $\tau$ to a choice expression $\{p(P)\}$ is the
conjunction of the formulas
$p({\bf t}) \lor \neg p({\bf t})$ over all tuples ${\bf t}$ in $[P]$.
Similarly, the result of applying $\tau$ to a choice expression $\{\tilde{p}(P)\}$
is the conjunction of the formulas $\tilde{p}({\bf t}) \lor \neg \tilde{p}({\bf t})$
over all tuples ${\bf t}$ in $[P]$.

For instance, the result of applying $\tau$ to rule $R_1$ (see
Section~\ref{rp}) is
\beq
\bigwedge_{1 \leq i,j \leq n} \left ( q(\overline i, \overline j) \lor \neg q(\overline i, \overline j)
\right ).
\eeq{tauR_1}

\subsection{Global Variables}\label{sem:global}

About a variable we say that it is {\sl global}
\begin{itemize}
\item
in a conditional literal $H : {\bf L}$, if it occurs in~$H$ but does not
occur in~${\bf L}$;
\item
in an aggregate literal $A$, $\no \; A$, or $\no \; \no A$, where $A$ is of one of
the forms~(\ref{ag1})--(\ref{ag3}), if it occurs in~$s, s_1,$ or $s_2$;
\item
in a rule~(\ref{rule3}), if it is global in at least one of the
expressions~$H_i$,~$B_j$;
\item
in a rule~(\ref{rule4}), if it occurs in $C$ or is global in at least one of
the expressions~~$B_j$.
\end{itemize}

An {\sl instance} of a rule~$R$ is any
rule that can be obtained from~$R$ by substituting
precomputed terms for all global variables.\footnote{\label{ft2} This
definition
differs slightly from that given by \citeANP{har14a}
\citeyear[Section~3.3]{har14a}. There,
substitutions that yield symbolic constants in the scope of arithmetical
operators do not form instances.
In a similar way, we treat variables in conditional literals
and aggregate literals (Sections~\ref{sem:cond} and~\ref{sem:ag}) differently
than how they are treated by \citeANP{har14a} \citeyear{har14a}.}
A literal or a rule is {\sl closed} if
it has no global variables. It is clear that any instance of a rule is closed.

For example, $X$ is global in the rule $R_2$ from Section~\ref{rp},
so that the instances of $R_2$ are rules of the form
$$
\ar \; r = \overline 1\;..\;\overline n \; \land \; \no \; {\it count}\{ Y : q(r,Y) \} = \overline 1
$$
for all precomputed terms $r$.
The variables $X$ and $Y$ are global in $R_4$; instances of $R_4$ are
$$
{\it d1}(r, s, r- s+\overline n) \; \ar \;  r = \overline 1\;..\;\overline n \; \land \;
 s = \overline 1\;..\;\overline n
$$
for all precomputed terms $r$ and $s$.
\subsection{Semantics of Conditional Literals}\label{sem:cond}

If~$t$ is a term,~${\bf x}$ is a tuple of distinct variables, and~${\bf r}$ is
a tuple of terms of the same length as~${\bf x}$, then the term obtained
from~$t$ by substituting~${\bf r}$
for~${\bf x}$ will be denoted by $t^{\bf x}_{\bf r}$.  Similar notation will be
used for the result of substituting~${\bf r}$ for~${\bf x}$ in expressions of
other kinds, such as literals and tuples of literals.

The result of applying $\tau$ to a closed conditional literal $H : {\bf L}$
is the conjunction of the formulas
$$\tau_{\lor} ({\bf L}^{\bf x}_{\bf r})\rar \tau_{\lor}(H^{\bf x}_{\bf r})$$
where ${\bf x}$ is the list of variables occurring in~$H : {\bf L}$,
over all tuples ${\bf r}$ of precomputed terms of the same length as~${\bf
x}$.

For instance, the result of applying $\tau$ to the arithmetic literal
$r = \overline 1 .. \overline n$, where~$r$ is a precomputed term, is
$\tau_{\lor}(\epsilon) \rar \tau_{\lor}(r = \overline 1 .. \overline n)$, where
$\epsilon$ is the tuple of length $0$.
The antecedent of this implication is $\top$. The consequent is~$\top$ if
$r$ is one of the numerals $\overline 1, \dots, \overline n$ and $\bot$ otherwise.

\subsection{Semantics of Aggregate Literals}\label{sem:ag}

In this section, the semantics of ground aggregates proposed by
\citeANP{fer05} \citeyear[Section~4.1]{fer05} is adapted to closed aggregate
literals.  Let~$E$
be a closed aggregate atom of one of the forms (\ref{ag1})--(\ref{ag3}), and
let~${\bf x}_i$ be the list of variables occurring in~${\bf t}_i:{\bf L}_i$
\hbox{($1\le i\le n$)}.  By~$A_i$ we denote the set of tuples~${\bf r}$ of
precomputed terms of the same length as~${\bf x}_i$.
By $A$ we denote the
set $\{(i, {\bf r}) : i \in \{1, \dots, n\}, {\bf r} \in A_i\}$.

Let $\Delta$ be a subset of $A$. Then by $[\Delta]$ we denote
the union of the sets $[({\bf t}_i)^{{\bf x}_i}_{\bf r}]$ for all pairs
$(i, {\bf r})\in\Delta$.
We say that $\Delta$ {\sl justifies}~$E$ with respect to a precomputed
term\footnote{This definition of the semantics of aggregates is
more complicated than that published in the original version of this document.
There, a set $\Delta$
either justifies an aggregate atom or not, without reference to a particular
precomputed term
$t$. The version here corrects a discrepancy between the semantics and the
behavior of {\sc gringo} in the case when $s$
represents a non-singleton set.\label{ft3a}
} $t$ if
\begin{itemize}
\item  $E$ is of the form (\ref{ag1}) and the relation $\prec$ holds
between $\widehat \alpha
[{\Delta}]$ and $t$,  or
\item $E$ is of the form (\ref{ag2}) and the relation $\prec$ holds between $t$ and
$\widehat \alpha [{\Delta}]$.
\end{itemize}
We say that $\Delta$ {\sl justifies}~$E$ with respect to a pair $t_1,t_2$ of precomputed
terms if
$E$ is of the
form (\ref{ag3}), the relation
$\prec_1$ holds between $t_1$ and $\widehat \alpha [\Delta]$, and the relation $\prec_2$ holds between $\widehat
\alpha [\Delta]$ and $t_2$.

If $t$ is a precomputed term, and $E$ is of form (\ref{ag1}) or (\ref{ag2}),
we define $\tau_t E$ as the conjunction of the implications
\beq
  \bigwedge_{(i, {\bf r})\in\Delta}\tau_{\lor}(({\bf L}_i)^{{\bf x}_i}_{\bf r})
  \,\rar\,
  \bigvee_{(i,{\bf r})\in A\setminus\Delta}\tau_{\lor}(({\bf L}_i)^{{\bf x}_i}_
  {\bf r})
\eeq{agdef}
over all sets $\Delta$ that do not justify~$E$ with respect to $t$.
If $t_1,t_2$ is a pair of precomputed terms, and $E$ is of form (\ref{ag3}),
we define $\tau_{t_1,t_2} E$ as the conjunction \eqref{agdef}
over all sets $\Delta$ that do not justify~$E$ with respect to $t_1,t_2$.

For instance, if $E$ is
${\it count}\{p(X):p(X)\} > 0$
then $\tau_{\o 0} E$ is the
(conjunction containing the single) implication expressing that $p(r)$ holds for at least one precomputed term $r$:
$$
\top \,\rar\, \bigvee_r p(r).
$$

For a closed aggregate atom $E$ of form \eqref{ag1} or \eqref{ag2},
\begin{itemize}
\item by $\tau E$ we denote the disjunction of formulas $\tau_t E$
over all terms $t$ in~$[s]$;
\item by $\tau (\no \; E)$ we denote the disjunction of formulas $\neg \tau_t E$
over all terms $t$ in~$[s]$; and
\item by $\tau (\no \; \no \; E)$ we denote the disjunction of formulas
$\neg \neg \tau_t E$ over all terms $t$ in~$[s]$.
\end{itemize}
It is clear that if $[s]$ is a singleton set $\{t\}$, then $\tau E$ is (the
disjunction containing only) $\tau_t E$.
For a closed aggregate atom $E$ of form \eqref{ag3},
\begin{itemize}
\item by $\tau E$ we denote the disjunction of formulas $\tau_{t_1,t_2} E$
over all pairs of precomputed terms $t_1, t_2$ such that $t_1$ in~$[s_1]$ and
$t_2$ in~$[s_2]$;
\item by $\tau(\no \; E)$ we denote the disjunction of formulas
$\neg \tau_{t_1,t_2} E$ over all pairs of precomputed terms $t_1, t_2$ such
that $t_1$ in~$[s_1]$ and $t_2$ in~$[s_2]$; and
\item by $\tau(\no \; \no \; E)$ we denote the disjunction of formulas
$\neg \neg \tau_{t_1,t_2} E$ over all pairs of precomputed terms $t_1, t_2$ such
that $t_1$ in~$[s_1]$ and $t_2$ in~$[s_2]$.
\end{itemize}

\subsection{Semantics of Rules and Programs} \label{ssec:grules}

For any rule~$R$ of form (\ref{rule3}), $\tau R$ stands for the set of the
formulas
$$
\tau B_1\land\cdots\land\tau B_m\rar\tau_{\land} H_1\lor\cdots\lor\tau_{\land} H_k
$$
for all instances~(\ref{rule3}) of~$R$.
For a rule of form (\ref{rule4}), $\tau R$ stands for the set of the
formulas
$$
\tau B_1\land\cdots\land\tau B_m\rar\tau C
$$
for all instances~(\ref{rule4}) of~$R$.
For any program~$\Pi$, $\tau \Pi$ stands for the union of the sets
$\tau R$ for all rules~$R$ of~$\Pi$.

A {\sl stable model} of a program~$\Pi$
is any stable model of~$\tau\Pi$ (in the sense of Section~\ref{sem:inf})
that does
not contain any pair of atoms of the form $p({\bf t})$, $\tilde{p}({\bf t})$.

\section{Simplifying $\tau\Pi$}

When we investigate the stable models of an AG program, it is often
useful to simplify the formulas obtained by applying transformation~$\tau$ to
its rules.  By simplifying an infinitary propositional formula we mean turning
it into a strongly equivalent formula that has
simpler syntactic structure.  The definition of strong equivalence,
introduced by~\citeANP{lif01} \citeyear{lif01}, is extended to infinitary
formulas by~\citeANP{har15a} \citeyear{har15a}.  Corollary~1 from that paper
shows that the stable models of an infinitary formula are not
affected by simplifying its parts.

Proofs of the theorems stated in this section are outlined
in the electronic appendix.

\subsection{Monotone and Anti-Monotone Aggregate Atoms}\label{sec:mon}

When a rule contains aggregate atoms, we can sometimes simplify the
implications~(\ref{agdef}) in the corresponding infinitary formula using
the theorems on monotone and anti-monotone aggregates from
\citeANP{har14a} \citeyear[Section~6.1]{har14a}.
The monotonicity or non-monotonicity of an
aggregate atom~(\ref{ag1}) can sometimes be established simply by looking
at its aggregate name~$\alpha$ and its relation symbol~$\prec$.  If~$\alpha$
is one of the symbols {\it count}, {\it sum+}, {\it max}, then~(\ref{ag1}) is
monotone when~$\prec$ is $<$ or $\leq$, and anti-monotone when~$\prec$ is
$>$ or $\geq$.  It is the other way around if~$\alpha$ is {\it min}.

Our semantics of aggregates is somewhat different from that adopted
by \citeANP{har14a} \citeyear[Section~3.5]{har14a}, as explained in
Footnote~\ref{ft3} (and also
in view of the difference in the treatment of variables discussed in
Footnote~\ref{ft2}, and the modification to the definition of ``justifies''
explained in Footnote~\ref{ft3a}).  Nevertheless, the statements and proofs of the two
theorems mentioned above remain essentially the same in the framework of AG. The
theorems show that the antecedent in (\ref{agdef}) can be dropped if $E$
is monotone, and that the consequent can be replaced by~$\bot$ if $E$ is
anti-monotone.  These simplifications produce strongly equivalent
formulas.

\subsection{Eliminating Equality from Aggregate Atoms}

If~$\prec$ in an aggregate atom~(\ref{ag1}) is $=$ then the following
theorem\footnote{The statement and proof of this theorem have been
modified with respect to the original version of this paper in accordance
with the change in the definition of ``justifies'' described in
Footnote~\ref{ft3a}.}  can be useful, in combination with the facts reviewed in
Section~\ref{sec:mon}:

\begin{theorem}\label{lem:equality}
If $E$ is a closed aggregate atom of the form
$$
\alpha\{{\bf t}_1 : {\bf L}_1\,;\ldots;\,{\bf t}_n : {\bf L}_n\}= s,
$$
$E_{\le}$ is
$$
\alpha\{{\bf t}_1 : {\bf L}_1\,;\ldots;\,{\bf t}_n : {\bf L}_n\}\leq s,
$$
and $E_{\ge}$ is
$$
\alpha\{{\bf t}_1 : {\bf L}_1\,;\ldots;\,{\bf t}_n : {\bf L}_n\}\geq s,
$$
then for any precomputed term $t$, $\tau_t E$ is strongly equivalent to $\tau_t E_{\le} \land
\tau_t E_{\ge}$.
\end{theorem}

\subsection{Properties of Counting} \label{sec:count}

For any set $S$, by $\lvert S \rvert$ we denote the cardinality of
$S$ if $S$ is finite, and $\infty$ otherwise.

\begin{theorem}\label{lem:geq}
For any closed aggregate atom~$E$ of the form
$$
{\it count}\{{\bf t}_1 : {\bf L}_1; \dots ; {\bf t}_n : {\bf L}_n \} \geq \overline m
$$
where $m$ is an integer and each ${\bf t}_i$ is interval-free, $\tau E$ is
strongly equivalent to
\beq
  \bigvee_{\Delta\subseteq A \atop \lvert [\Delta] \rvert = m}
  \bigwedge_{(i,{\bf r}) \in \Delta}
  \tau_{\lor}(({\bf L}_i)^{{\bf x}_i}_{\bf r}).
\eeq{trans_geq}
\end{theorem}

\begin{theorem}\label{lem:leq}
For any closed aggregate atom $E$ of the form
$$
{\it count}\{{\bf t}_1 : {\bf L}_1; \dots ; {\bf t}_n : {\bf L}_n \} \leq \overline m
$$
where $m$ is an integer and each ${\bf t}_i$ is interval-free, $\tau E$ is
strongly equivalent to
\beq
  \bigwedge_{\Delta \subseteq A \atop \lvert [\Delta] \rvert = m+1}  \neg
  \bigwedge_{(i, {\bf r}) \in\Delta} \tau_{\lor}
  (({\bf L}_i)^{{\bf x}_i}_{\bf r}).
\eeq{trans_leq}
\end{theorem}

Without the assumption that each ${\bf t}_i$ is interval-free the assertions
of the theorems would be incorrect. For instance, if $E$ is
${\it count}\{\overline 1..\overline 2:p\} \geq \overline 1$ then $\tau E$ is $\top \rar p$, and
(\ref{trans_geq}) is~$\bot$.

In the special (but common) case when $E$ has the form
${\it count}\{{\bf x}: {\bf L}\} \geq \overline m$,
where ${\bf x}$ is a tuple of variables and each variable occurring in
${\bf L}$ occurs also in ${\bf x}$,
the condition $\lvert [\Delta] \rvert = m$ in (\ref{trans_geq})
can be replaced by $\lvert \Delta \rvert = m$. Indeed, in this case
$\Delta$ and $[\Delta]$ have the same cardinality because $[\Delta]$
is the set of tuples ${\bf r}$ of terms such that $(1, {\bf r}) \in \Delta$.
Similarly,  the condition $\lvert [\Delta] \rvert = m+1$ in (\ref{trans_leq})
can be replaced by $\lvert \Delta \rvert = m+1$ if
$E$ has the form ${\it count}\{{\bf x}: {\bf L}\} \leq \overline m$.

\section{Conclusion}

We proposed a definition of stable models for programs in the language AG and
stated a few theorems that facilitate reasoning about them.  This definition
can be viewed as a specification for the answer set
system {\sc clingo} (see Footnote~\ref{ft1}) and other systems with the
same input language.  If such a system terminates given the ASCII
representation of an AG program~$\Pi$ as input, and produces neither error
messages nor warnings, then its output is expected to represent the stable
models of~$\Pi$.

\section*{Acknowledgements}

We are grateful to the anonymous referees for useful comments.

\newpage


\begin{thebibliography}{}

\bibitem[\protect\citeauthoryear{Calimeri, Faber, Gebser, Ianni, Kaminski,
  Krennwallner, Leone, Ricca, and Schaub}{Calimeri
  et~al\mbox{.}}{2012}]{aspcore2}
{\sc Calimeri, F.}, {\sc Faber, W.}, {\sc Gebser, M.}, {\sc Ianni, G.}, {\sc
  Kaminski, R.}, {\sc Krennwallner, T.}, {\sc Leone, N.}, {\sc Ricca, F.}, {\sc
  and} {\sc Schaub, T.} 2012.
\newblock {ASP-Core-2}: Input language format.
\newblock Available at
  \url{https://www.mat.unical.it/aspcomp2013/files/ASP-CORE-2.0.pdf}.

\bibitem[\protect\citeauthoryear{Ferraris}{Ferraris}{2005}]{fer05}
{\sc Ferraris, P.} 2005.
\newblock Answer sets for propositional theories.
\newblock In {\em Proceedings of International Conference on Logic Programming
  and Nonmonotonic Reasoning ({LPNMR})}. 119--131.

\bibitem[\protect\citeauthoryear{Ferraris and Lifschitz}{Ferraris and
  Lifschitz}{2005}]{fer05b}
{\sc Ferraris, P.} {\sc and} {\sc Lifschitz, V.} 2005.
\newblock Weight constraints as nested expressions.
\newblock {\em Theory and Practice of Logic Programming\/}~{\em 5,\/}~1--2,
  45--74.

\bibitem[\protect\citeauthoryear{Gebser, Kaminski, Kaufmann, and Schaub}{Gebser
  et~al\mbox{.}}{2011}]{geb11b}
{\sc Gebser, M.}, {\sc Kaminski, R.}, {\sc Kaufmann, B.}, {\sc and} {\sc
  Schaub, T.} 2011.
\newblock Challenges in answer set solving.
\newblock In {\em Logic programming, knowledge representation, and nonmonotonic
  reasoning}. Springer, 74--90.

\bibitem[\protect\citeauthoryear{Harrison, Lifschitz, Pearce, and
  Valverde}{Harrison et~al\mbox{.}}{2015}]{har15a}
{\sc Harrison, A.}, {\sc Lifschitz, V.}, {\sc Pearce, D.}, {\sc and} {\sc
  Valverde, A.} 2015.
\newblock Infinitary equilibrium logic and strong equivalence.
\newblock In {\em Proceedings of International Conference on Logic Programming
  and Nonmonotonic Reasoning ({LPNMR})}.
\newblock {\tt http://www.cs.utexas.edu/users/vl/ papers/iel\_lpnmr.pdf}; to
  appear.

\bibitem[\protect\citeauthoryear{Harrison, Lifschitz, and Yang}{Harrison
  et~al\mbox{.}}{2014}]{har14a}
{\sc Harrison, A.}, {\sc Lifschitz, V.}, {\sc and} {\sc Yang, F.} 2014.
\newblock The semantics of {G}ringo and infinitary propositional formulas.
\newblock In {\em Proceedings of International Conference on Principles of
  Knowledge Representation and Reasoning (KR)}.

\bibitem[\protect\citeauthoryear{Lifschitz, Pearce, and Valverde}{Lifschitz
  et~al\mbox{.}}{2001}]{lif01}
{\sc Lifschitz, V.}, {\sc Pearce, D.}, {\sc and} {\sc Valverde, A.} 2001.
\newblock Strongly equivalent logic programs.
\newblock {\em ACM Transactions on Computational Logic\/}~{\em 2}, 526--541.

\bibitem[\protect\citeauthoryear{Lifschitz, Tang, and Turner}{Lifschitz
  et~al\mbox{.}}{1999}]{lif99d}
{\sc Lifschitz, V.}, {\sc Tang, L.~R.}, {\sc and} {\sc Turner, H.} 1999.
\newblock Nested expressions in logic programs.
\newblock {\em Annals of Mathematics and Artificial Intelligence\/}~{\em 25},
  369--389.

\bibitem[\protect\citeauthoryear{Truszczynski}{Truszczynski}{2012}]{tru12}
{\sc Truszczynski, M.} 2012.
\newblock Connecting first-order {ASP} and the logic {FO(ID)} through reducts.
\newblock In {\em Correct Reasoning: Essays on Logic-Based AI in Honor of
  Vladimir Lifschitz}, {E.~Erdem}, {J.~Lee}, {Y.~Lierler}, {and} {D.~Pearce},
  Eds. Springer, 543--559.

\end{thebibliography}

\end{document}


\appendix
\pagenumbering{gobble}
\title[Online appendix]{{\large\textnormal{Online appendix for the paper}}   \\
Abstract Gringo
\\
{\large\textnormal{published in Theory and Practice of Logic Programming}}
}
\author[Gebser, Harrison, Kaminski, Lifschitz, and Schaub]
{MARTIN GEBSER\\
Aalto University, HIIT, Finland\\
University of Potsdam, Germany\\
gebser@cs.uni-potsdam.de\\
\and
AMELIA HARRISON\\
Univeristy of Texas at Austin, USA\\
ameliaj@cs.utexas.edu\\
\and
ROLAND KAMINSKI\\
University of Potsdam, Germany\\
kaminski@cs.uni-potsdam.de\\
\and
VLADIMIR LIFSCHITZ\\
Univeristy of Texas at Austin, USA\\
vl@cs.utexas.edu\\
\and
TORSTEN SCHAUB\\
University of Potsdam, Germany\\
INRIA Rennes, France\\
torsten@cs.uni-potsdam.de\\
}


\makeatletter
{%
  \newpage
  \begingroup
    \newpage
    \global\@topnum\z@
    \titlefntrue
    \def\thefootnote{\@fnsymbol\c@footnote}%
    \def\@makefnmark{\hbox{$\@thefnmark$}}%
    \thispagestyle{titlepage}\@thanks
  \endgroup
  \vspace*{-15\p@}%
  {\centering \sloppy
   \pe@rl{#1}%
   {\normalfont\LARGE\itshape \@title\par}%
   \vskip 16\p@
   {\normalfont\normalsize\rmfamily
    \let\authorbreak\auth@rbreak
    \let\and\@nd
    \b@at
      \@author
    \end{author@tabular}%
   \par}%
  }%
}
\makeatother

\section*{Proofs of Theorems~1--3}
\label{app:proofs}

\begin{proof*}[Proof of Theorem~1]
We will show that for any term $t$, the set of conjunctive terms of
$\tau_t E$ is the union
of the sets of conjunctive terms of $\tau_t E_{\leq}$ and $\tau_t E_{\geq}$.
For any subset $\Delta$ of~$A$,
\begin{center}
\begin{tabular}{l  l}
    & (22) is a conjunctive term of $\tau_t E$ \\
iff & $\Delta$ does not justify $E$ with respect to $t$ \\
iff & $\widehat{\alpha}[\Delta] \not = t$ \\
iff & $\widehat{\alpha}[\Delta] < t$ or $\widehat{\alpha}[\Delta] > t$ \\
iff & $\Delta$ does not justify $E_{\ge}$ with respect to $t$ or $\Delta$ does not justify
$E_{\le}$ with respect to $t$\\
iff & (22) is a conjunctive term of $\tau_t E_{\ge}$ or of
$\tau_t E_{\le}$\\
iff & (22) is a conjunctive term of $\tau_t E_{\le} \land \tau_t E_{\ge}.$
\end{tabular}
\end{center}
\end{proof*}

\newpage

\begin{proof*}[Proof of Theorem~2]
Since $[\o m]$ is the singleton set $\{\o m\}$,
$\tau E$ is $\tau_{\o m} E$.
Since~$E$ is monotone,
the antecedent of~(22) can be dropped
(Section~5.1), so that $\tau_{\o m} E$ is strongly equivalent to
\[
  \bigwedge_{ \Delta \subseteq A \atop \lvert [\Delta] \rvert < m}
  \bigvee_{(i,{\bf r}) \in A\setminus\Delta}
  \tau_{\lor}(({\bf L}_i)^{{\bf x}_i}_{\bf r}).
\tag{25}
\label{mono_geq}
\]
To derive (25) from (23) in HT\/$^\infty$,
assume (23). We will reason by cases, with one case
corresponding to each disjunctive term
\[
  \bigwedge_{(i,{\bf r}) \in \Delta}
  \tau_{\lor}(({\bf L}_i)^{{\bf x}_i}_{\bf r})
\tag{26}
\label{dterm}
\]
of (23).
Let $\Delta'$ be a subset of $A$ such that
$\lvert [\Delta']\rvert < m$.
We will show that the conjunctive term
of (25) corresponding to $\Delta'$ can be derived from
(26).  Since
\[
\lvert [\Delta'] \rvert < m = \lvert [\Delta]\rvert,
\tag{27}
\label{ineq}
\]
there exists a pair $(i, {\bf r})$ that is an element of $\Delta$ but
not an element of $\Delta'$. Indeed, if \hbox{$\Delta \subseteq \Delta'$} then
$[\Delta] \subseteq [\Delta']$, which contradicts (27).
Since $(i, {\bf r}) \in \Delta$, from~(26)
we can derive $\tau_{\lor}(({\bf L}_i)_{\bf r}^{{\bf x}_i})$. Since
$(i, {\bf r}) \in A \setminus \Delta'$, we can further derive
$$
 \bigvee_{(i,{\bf r}) \in A\setminus\Delta'}
 \tau_{\lor}(({\bf L}_i)^{{\bf x}_i}_{\bf r}).
$$
It follows that each conjunctive term of (25)
can be derived from (26).

We will prove by induction on $m$ that
(23) can be derived from (25) in HT\/$^\infty$.
Base case: when $m = 0$ the disjunctive term of (23)
corresponding to the empty $\Delta$ is $\top$.
Inductive step: assume that (23) can be derived from
(25), and assume
\[
  \bigwedge_{\Delta \subseteq A \atop \lvert [\Delta]\rvert < m + 1}
  \bigvee_{(i,{\bf r}) \in A\setminus\Delta}
  \tau_{\lor}(({\bf L}_i)^{{\bf x}_i}_{\bf r}).
\tag{28}
\label{monok}
\]
From (28) we can derive (25), and consequently
(23).
Now we reason by cases, with one case corresponding to each disjunctive
term of (23).
Assume
\[
  \bigwedge_{(i,{\bf r}) \in \Sigma}
  \tau_{\lor}(({\bf L}_i)^{{\bf x}_i}_{\bf r})
\tag{29}
\label{dterm2}
\]
where $\Sigma$ is a subset of $A$ such that $\lvert [\Sigma]\rvert = m$.
Consider the set
$$\Sigma' = \{(i, {\bf r}) : [({\bf t}_i)^{{\bf x}_i}_{\bf r}]
\subseteq [\Sigma]\}.$$
By the definition of $[\Sigma]$, for any $(i, {\bf r}) \in \Sigma$,
$[({\bf t}_i)^{{\bf x}_i}_{\bf r}] \subseteq [\Sigma]$. So $\Sigma \subseteq
\Sigma'$. It follows that $[\Sigma] \subseteq [\Sigma']$.
On the other hand,
$$[\Sigma'] =
\bigcup_{(i, {\bf r}) \in \Sigma'} [({\bf t}_i)^{{\bf x}_i}_{\bf r}] =
\bigcup_{(i, {\bf r})\;:\;
[({\bf t}_i)^{{\bf x}_i}_{\bf r}] \subseteq [\Sigma]}
[({\bf t}_i)^{{\bf x}_i}_{{\bf r}}] \subseteq [\Sigma].$$
Consequently $[\Sigma] = [\Sigma']$, and
$\lvert [\Sigma'] \rvert = \lvert [\Sigma] \rvert = m.$
From (28),
\[
  \bigvee_{(i,{\bf r}) \in A\setminus\Sigma'}
  \tau_{\lor}(({\bf L}_i)^{{\bf x}_i}_{\bf r}).
\tag{30}
\label{monok_res}
\]
Again, we reason by cases, with one case corresponding to each disjunctive
term of (30). Assume $\tau_{\lor}(({\bf L}_j)^
{{\bf x}_j}_{\bf s})$, where $(j, {\bf s}) \in A \setminus \Sigma'$.
Combining assumption (29) and $\tau_{\lor}(({\bf L}_j)^
{{\bf x}_j}_{\bf s})$, we derive
\[
  \bigwedge_{(i, {\bf r}) \in \Sigma \cup \{(j,{\bf s})\}}
  \tau_{\lor}(({\bf L}_i)^{{\bf x}_i}_{\bf r}).
\tag{31}
\label{combin}
\]
Consider the set $[\Sigma \cup \{(j, {\bf s})\}]$, that is,
\[
[\Sigma] \cup [({\bf t}_j)^{{\bf x}_j}_{\bf s}].
\tag{32}
\label{31}
\]
Recall that the cardinality
of $[\Sigma]$ is $m$. Since ${\bf t}_j$ is interval-free,
the cardinality of $[({\bf t}_j)^{{\bf x}_j}_{\bf s}]$ is at most $1$.
Furthermore,
since $(j, {\bf s}) \not \in \Sigma'$ it follows that
$$[({\bf t}_j)^{{\bf x}_j}_{\bf s}] \not \subseteq [\Sigma],$$
so that $[({\bf t}_j)^{{\bf x}_j}_{\bf s}]$ is nonempty.
Consequently,
the set is a singleton,
and therefore $[\Sigma]$ is disjoint from it.
It follows that the cardinality of (\ref{31}) is $m+1$.
So from (\ref{combin}) we can derive
$$
  \bigvee_{\Delta\subseteq A \atop \lvert [\Delta] \rvert = m+1}
  \bigwedge_{(i,{\bf r}) \in \Delta}
  \tau_{\lor}(({\bf L}_i)^{{\bf x}_i}_{\bf r}).
$$
\end{proof*}

\begin{proof*}[Proof of Theorem~3]
Since $[\o m]$ is the singleton set $\{\o m\}$,
$\tau E$ is $\tau_{\o m} E$.
Since the
consequent of~(22) can be replaced in this case by~$\bot$,
$\tau_{\o m} E$ is strongly equivalent to
\[
  \bigwedge_{\Delta \subseteq A \atop \lvert [\Delta] \rvert > m}
  \neg \bigwedge_{(i, {\bf r}) \in\Delta}
  \tau_{\lor}(({\bf L}_i)^{{\bf x}_i}_{\bf r}).
\tag{33}
\label{trans}
\]
Every conjunctive term of (24) is a conjunctive term of
(\ref{trans}). To derive (\ref{trans}) from (24), consider a
set $\Delta$ such that $\lvert [\Delta] \rvert > m$.
Let $f(i,{\bf r})$ stand for the set
$[({\bf t}_i)^{{\bf x}_i}_{\bf r}]$. Since each ${\bf t}_i$ is
interval-free, this set is either empty or a singleton.
Let ${\bf s}_1,\dots,{\bf s}_{m+1}$
be $m+1$ distinct elements of $[\Delta]$.
Choose elements $(i_1,{\bf r}_1),\dots,(i_{m+1},{\bf r}_{m+1})$ of
$\Delta$ such that each $s_k$ belongs to $f(i_k,{\bf r}_k)$, and let
$\Delta'$ be $\{(i_1,{\bf r}_1),\dots,(i_{m+1},{\bf r}_{m+1})\}$.
The cardinality of $[\Delta']$ is at least $m+1$, because this set
includes ${\bf s}_1,\dots,{\bf s}_{m+1}$.  On the other hand, it is at
most $m+1$, because this set is the union of $m+1$ sets of cardinality at
most $1$.
Consequently, $\lvert [\Delta']\rvert = m+1$.  From (24)
we can conclude in HT\/$^\infty$ that
\[
  \neg \bigwedge_{(i, {\bf r}) \in\Delta'} \tau_{\lor}
  (({\bf L}_i)^{{\bf x}_i}_{\bf r}).
\tag{34}
\label{cterm}
\]
Then the conjunctive term
$$\neg \bigwedge_{(i, {\bf r}) \in\Delta}
  \tau_{\lor}(({\bf L}_i)^{{\bf x}_i}_{\bf r})$$
of~(\ref{trans}) follows, because $\Delta' \subseteq \Delta$.\end{proof*}

\section*{Correctness of the $n$-Queens Program} \label{sec:nqueens}

In this section, we prove the correctness of the program $K$,
consisting of rules $R_1,\dots,R_7$ (Sections~2.3 and~3).

The $n$-queens problem involves placing $n$ queens on an $n \times n$
chess board such that no two queens threaten each other. We will represent
squares by pairs of integers $(i, j)$ where
\hbox{$1 \leq i,j \leq n$}. Two squares $(i_1, j_1)$ and $(i_2, j_2)$ are said
to be in the same row if $i_1 = i_2$; in the same column
if $j_1 = j_2$; and in the same diagonal if $\lvert i_1 - i_2 \rvert =
\lvert j_1 - j_2 \rvert$.
A set $Q$ of $n$ squares is a {\sl solution} to the $n$-queens problem if
no two elements of $Q$ are in the same row,
in the same column,
or in the same diagonal.

For any stable model $I$ of $K$, by $Q_I$ we denote the set of
pairs $(i, j)$ such that $q(\o i, \o j) \in I$.

\medskip

\noindent\emph{Theorem 4}\\
For each stable model $I$ of $K$, $Q_I$ is a
solution to the $n$-queens problem.  Furthermore, for each solution $Q$ to
the $n$-queens problem there is exactly one stable model $I$ of $K$ such that
$Q_I = Q$.

\medskip

\subsection*{Review: Supported Models and Constraints}

We start by reviewing two familiar facts that will be
useful in proving Theorem~4.

An {\sl infinitary program} is a conjunction of (possibly infinitely
many) infinitary formulas of the form $G \rar A$, where $A$ is an atom.
We say that an interpretation $I$ is
{\sl supported} by an infinitary program $\Pi$ if each atom $A$ from $I$ is
the consequent of a conjunctive term~$G\rar A$ of $\Pi$ such that $I\models G$.
\citeANP{lif13a} \citeyear{lif13a} give a condition, ``tightness on an
interpretation,''  under which the stable models of
an infinitary program are identical to its supported models.
Proposition~\ref{thm:etight} below
gives a simpler condition of this kind that is sufficient for our purposes.

We say that an atom $A$ {\sl occurs nonnegated in a formula $F$} if
\begin{itemize}
\item $F$ is $A$, or
\item $F$ is of the form $\mathcal{H}^\land$ or $\mathcal{H}^\lor$ and $A$ occurs nonnegated in at least one element of $\mathcal{H}$, or
\item $F$ is of the form $G \rar H$, where $H$ is different from $\bot$, and
$A$ occurs nonnegated in $G$ or in $H$.
\end{itemize}
It is clear, for instance, that no atom occurs nonnegated in a formula of the
form $\neg F$.

The {\sl positive dependency graph} of an infinitary program $\Pi$ is the
directed graph containing a vertex for each atom occuring in $\Pi$, and an
edge from $A$ to $B$ for every conjunctive term $G \rar A$ of $\Pi$ and every
atom $B$ that occurs nonnegated in $G$.
We say that an infinitary program $\Pi$ is {\sl extratight}
if the positive dependency graph of $\Pi$ contains no infinite paths.

The following fact is immediate from \cite[Lemma 2]{lif13a}.

\begin{proposition}\label{thm:etight}
For any model $I$ of an extratight infinitary program $\Pi$, $I$ is
stable iff $I$ is supported by~$\Pi$.
\end{proposition}

A {\sl constraint} is an infinitary formula of the form $\neg F$
(which is shorthand for $F\rar\bot$). The
following theorem is a straightforward generalization of
Proposition 4  from \cite{fer05e}.

\begin{proposition}\label{thm:cons}
Let $\mathcal{H}_1$  be a set of infinitary formulas and $\mathcal{H}_2$ be
a set of constraints. A set $I$ of
atoms  is a stable model of $\mathcal{H}_1 \cup \mathcal{H}_2$ iff $I$ is a
stable model of $\mathcal{H}_1$ and satisfies all formulas in~$\mathcal{H}_2$.
\end{proposition}

\begin{proof*}
{\em Case 1:} Every formula in $\mathcal{H}_1 \cup \mathcal{H}_2$ is satisfied
by $I$. For each formula $\neg F$ in $\mathcal{H}_2$, $I$ does not satisfy $F$.
So the reduct of each formula in $\mathcal{H}_2$ w.r.t. $I$ is $\neg \bot$.
It follows that the set of reducts of all formulas in
$\mathcal{H}_1 \cup \mathcal{H}_2$
is satisfied by the same interpretations as the set of reducts of all
formulas in $\mathcal{H}_1$.  Consequently, $I$ is minimal among the sets
satisfying the reducts of all formulas from
$\mathcal{H}_1 \cup \mathcal{H}_2$ iff it is minimal among the sets
satisfying the reducts of all formulas from $\mathcal{H}_1$. {\em Case~2: }
Some formula $F$ in $\mathcal{H}_1 \cup \mathcal{H}_2$ is not satisfied by $I$.
Then $I$ is not a stable model of $\mathcal{H}_1 \cup \mathcal{H}_2$.
If $F \in \mathcal{H}_1$ then $I$ is not a stable model of $\mathcal{H}_1$.
Otherwise, it is not true that $I$ satisfies all formulas in $\mathcal{H}_2$.
\end{proof*}

\subsection*{Proof of Theorem 4}

To simplify notation, we will identify each set $Q$ of squares with the set of
atoms $q(\o i, \o j)$ where \hbox{$(i, j) \in Q$}. By $D_n$ we denote the set
of atoms of the forms $\i{d1}(\o i, \o j, \o{i-j+n})$ and $\i{d2}(\o i, \o j,
\o{i+j-1})$ for all $i,j$ from $\{1,\dots,n\}$. Recall that the rules
of the program $K$ are denoted by $R_1,\dots,R_7$.

\begin{lemma}\label{lem:sup}
A set of atoms is a stable model of
\[
\tau R_1 \cup \tau R_4 \cup \tau R_5
\tag{35}
\label{supset}
\]
iff it is of the form $Q \cup D_n$ where $Q$ is a set of squares.
\end{lemma}

\begin{proof*}
We can turn (\ref{supset}) into a strongly
equivalent infinitary program as follows. The result of applying $\tau$ to
$R_1$ is (21).
Each conjunctive term in this formula is strongly equivalent to
\[
\neg \neg q(\o i, \o j) \rar q(\o i, \o j).
\tag{36}
\label{etightR_1}
\]
The set $\tau R_4$ is strongly equivalent to
the set of formulas
\[
  \top \rar d1(\o i,\o j,\o{i-j+n})
\tag{37}
\label{r4}
\]
($1 \leq i,j \leq n$).
(We take into account that $\tau(\o i = \o 1..\o n)$ is equivalent to
$\top$ if $1 \leq i \leq n$ and to~$\bot$ otherwise, and similarly for $j$.)
Similarly, $\tau R_5$ is strongly equivalent to the set of formulas
\[
  \top \rar d2(\o i,\o j,\o{i+j-1})
\tag{38}
\label{r5}
\]
$(1 \leq i,j \leq n)$.
Consequently, (\ref{supset}) is strongly equivalent to the conjunction $H$ of
formulas (\ref{etightR_1})--(\ref{r5}).
It is easy to check that $H$ is an extratight infinitary program, so that
by Proposition~\ref{thm:etight} its stable models are identical to its
supported models.
A set $I$ of atoms is a model of $H$ iff $D_n \subseteq I$. Furthermore,
$I$ is supported iff every element of $I$ has the form $q(\o i, \o j)$ or
is an element of $D_n$. Consequently, supported models of $H$ are sets of the
form $Q \cup D_n$ where $Q$ is a set of squares.
\end{proof*}

\begin{lemma}\label{lem:main}
A set $I$ of atoms is a stable model of $\tau K$ iff
it has the form $Q \cup D_n$, where $Q$ is a solution to the $n$-queens
problem.
\end{lemma}

\begin{proof*}
Let $\mathcal{H}_1$ be (\ref{supset}) and $\mathcal{H}_2$ be
$$
\tau R_2 \cup \tau R_3 \cup \tau R_6 \cup \tau R_7.
$$
All formulas in $\mathcal{H}_2$ are constraints. Consequently,
by Proposition~\ref{thm:cons}, $I$ is a stable
model of $\tau K$ iff it is a stable model of $\mathcal{H}_1$ and satisfies all formulas in $\mathcal{H}_2$.
By Lemma~\ref{lem:sup}, $I$ is a stable model of $\mathcal{H}_1$ iff it is
of the form $Q \cup D_n$, where $Q$ is a set of squares. It remains to show
that a set $I$ of the form $Q \cup D_n$ satisfies all formulas in
$\mathcal{H}_2$ iff $Q$ is a solution to the $n$-queens problem.
Specifically, we will show that for any set $I$ of the form $Q \cup D_n$
\begin{itemize}
\item[(i)] $I$ satisfies $\tau R_2$ iff for all $i \in \{1,\dots,n\}$, $I$
contains exactly one atom of the form $q(\o i, \o j)$;
\item[(ii)] $I$ satisfies $\tau R_3$ iff~for all $j \in \{1,\dots,n\}$, $I$
contains exactly one atom of the form $q(\o i, \o j)$;
\item[(iii)] $I$ satisfies $\tau R_6 \cup \tau R_7$ iff no two squares in $I$
are in the same diagonal.
\end{itemize}

To prove (i), note first that $\tau R_2$ is equivalent to  the set of formulas
$$
\neg \neg \left (\tau(\i{count}\{Y: q(\o i,Y)\} = \o 1) \right )
$$
$(1 \leq i \leq n)$. Let $E$ be the aggregate atom above. Since $[\o 1]$ is a
singleton set, $\tau E$ is the same as $\tau_{\o 1} E$. By Theorem~1, this set is strongly
equivalent to the set of formulas
\[
\neg \neg \left (\tau_{\o 1}(\i{count}\{Y:
  q(\o i,Y)\} \leq \o 1) \land \tau_{\o 1}(\i{count}\{Y: q(\o i,Y)\} \geq \o 1)
  \right ).
\tag{39}
\label{r2expand}
\]
Again note that the result of applying $\tau$ to first aggregate atom in \eqref{r2expand}
is the same as the result of applying $\tau_{\o 1}$.
Then by Theorem~3 and the comment at the end of Section~5.3,
$\tau$ applied to this
aggregate atom is strongly equivalent to
$$
\bigwedge_{\Delta \subseteq A \atop \lvert \Delta \rvert = 2} \neg
  \bigwedge_{(1, r) \in \Delta} q(\o i, r).
$$
This formula can be written as
$$
\bigwedge_{\Sigma \subseteq P \atop \lvert \Sigma \rvert = 2} \neg
  \bigwedge_{r \in \Sigma} q(\o i, r),
$$
where $P$ is the set of precomputed terms.
It is easy to see that $I$ satisfies this formula iff it contains at most
one atom of the form  $q(\o i,r)$. On the other hand, by Theorem~2,
the result of applying $\tau$ to the second aggregate atom in (\ref{r2expand})
is strongly equivalent to
$$
  \bigvee_{\Delta \subseteq A \atop \lvert \Delta \rvert = 1}
  \bigwedge_{(1, r) \in\Delta} q(\o i, r).
$$
Similar reasoning shows that $I$ satisfies this formula iff it
contains at least
one atom of the form  $q(\o i,r)$. Since $I = Q \cup D_n$,
$r$ in this atom is one of $\o 1,\dots, \o n$.

Claim (ii) is proved in a similar way.

To prove (iii), note first that two squares
$(\o{i_1}, \o{j_1}), (\o{i_2}, \o{j_2})$ are in the same diagonal iff
there exists a $k \in \{1, \dots, 2n-1\}$ such that
\[
d1(\o i_1, \o j_1, \o k), d1(\o i_2, \o j_2, \o k) \in D_n
\tag{40}
\label{c1}
\]
or
\[
d2(\o i_1, \o j_1, \o k), d2(\o i_2, \o j_2, \o k) \in D_n.
\tag{41}
\label{c2}
\]
We will show that a set $I$ of the form $Q \cup D_n$ does not satisfy
$\tau R_6$ iff
there exists a $k$ such that~(\ref{c1}) holds for two distinct elements
$q(\o{i_1}, \o{j_1}), q(\o{i_2}, \o{j_2}) \in Q$, and that it does not
satisfy
$\tau R_7$ iff there exists a $k$ such that (\ref{c2}) holds for such two
elements.  The result of applying $\tau$ to $R_6$ is strongly
equivalent to the set of formulas
\[
  \neg \tau(2 \leq \i{count}\{\o 0, q(X,Y) : q(X,Y), d1(X,Y,\o k)\})
\tag{42}
\label{tauR_6}
\]
$(1 \leq k \leq 2n-1)$. Formula (\ref{tauR_6}) is identical to
$$
\neg \tau ( \i{count}\{X,Y : q(X,Y), d1(X,Y,\o k)\} \geq 2 ).
$$
In view of Theorem~2, it follows that it is strongly equivalent
to
$$
\neg \bigvee_{\Delta \subseteq A \atop \lvert \Delta
 \rvert = 2} \bigwedge_{(1, (r,s)) \in \Delta} (q(r, s) \land d1(r, s,  \o k))
$$
$(1 \leq k \leq 2n-1)$. This formula can be written as
\[
 \neg \bigvee_{\Sigma \subseteq P \times P \atop \lvert \Sigma
 \rvert = 2} \bigwedge_{(r,s) \in \Sigma} (q(r, s) \land d1(r, s,  \o k)).
\tag{43}
\label{r6simp}\]
For any set $Q$ of squares,
\begin{center}
\begin{tabular}{l  l}
    & $Q \cup D_n$ does not satisfy (\ref{r6simp}) \\
iff & there exist two distinct pairs $(r_1, s_1), (r_2, s_2)$ from
    $P \times P$ such that\\
    & $q(r_1,s_1),q(r_2, s_2) \in Q$ and $d1(r_1,s_1,\o k),
    d1(r_2, s_2, \o k) \in D_n$\\
iff & there exist two distinct squares $(\o i_1, \o j_1), (\o i_2, \o j_2) \in
    Q$ such that (\ref{c1}) holds.
\end{tabular}
\end{center}

The claim about (\ref{c2}) is proved in a similar way.
\end{proof*}

Theorem~4 is immediate from the lemma.
